\documentclass[twocolumn, secnumarabic, amssymb, amsmath,nobibnotes, aps, prl]{revtex4}

\begin{document}

\title{Thermodynamics of An Ideal Generalized Gas: I Thermodynamic Laws} 
\author{B. H. Lavenda}
\email{bernard.lavenda@unicam.it}
\affiliation{Universit\'a degli Studi, Camerino 62032 (MC) Italy}
\date{\today}
\newcommand{\sumi}{\sum_{i=1}^{n}\,}
\newcommand{\sumj}{\sum_{j=1}^{n}\,}
\newcommand{\sumk}{\sum_{i=1}^{k}\,}
\newcommand{\sumgr}{\sum_{i=1}^{>}\,}
\newcommand{\sumle}{\sum_{i=1}^{\le}\,}
\newcommand{\prodi}{\prod_{i=1}^n\,}
\newcommand{\half}{\mbox{\small{$\frac{1}{2}$}}}
\newcommand{\third}{\mbox{\small{$\frac{1}{3}$}}}\newcommand{\twothirds}{\mbox{\small{$\frac{2}{3}$}}}

\begin{abstract}
The equations of state for an ideal generalized gas, like an ideal quantum gas, are expressed in terms of power laws of the temperature. The reduction of an ideal generalized gas to an ideal classical case occurs when the characteristic empirical temperature exponents in the thermal equation of state and in the absolute temperature coincide in contrast to the merger of an ideal quantum gas with an ideal classical gas in the high temperature limit. A corollary to Carnot's theorem is proved asserting that the ratio of the work done over a cycle to the heat absorbed to increase the temperature at constant volume is the same for all bodies at the same volume. As power means, the energy and entropy are incomparable and a new adiabatic potential is introduced by showing that the volume raised to a characteristic exponent is also the integrating factor for the quantity of heat so that the second law can be based on the property that power means are monotonically increasing functions of their order.
\end{abstract}
\maketitle
\section{incomparable thermodynamic laws}
Thermodynamics distinguishes itself on being able to take into account processes involving heat transfer. According to J. J. Thomson \cite{JJ}, heat is the \lq uncontrollable\rq\ form of work, and temperature is its measure. J. P. Joule appreciated that heat and work were interconvertible, depending only on a constant for the units chosen to measure heat and mechanical work.\par
The first law was formulated as an expression for the conservation of internal energy even in the presence of heat transfer. The second law placed limitations on the amount of heat that could be converted into mechanical work. According to William Thomson (Lord Kelvin) \cite{Kelvin}
\begin{quote}
it is impossible to construct an engine which when operated in a cycle will produce no effect other than the extraction of heat from a reservoir and the performance of an equivalent amount of work.
\end{quote}
The second law also succinctly sums up observations of nature like \lq\lq heat always passes from hotter to colder, and never in the reverse direction\rq\rq\cite{Caratheodory}, at constant volume. It also asserts that \lq\lq heat will be absorbed as a gas expands to keep its temperature constant\rq\rq.\par
Nicolas-L\'eonard-Sadi Carnot \cite{Carnot} not only discovered that there is an upper bound to the efficiency of engines operating in a closed cycle, but, also called attention to a truly mathematical invariant quantity that inspired  Thomson's later researches into the discovery of that entity. This invariant would be the same for all susbstances at the same temperature, and Thomson, equated it with  the temperature measured on the ideal gas absolute scale. This factor proved to be an integrating denominator for the quantity of heat, and became a matter of contention between R. Clausius and Thomson for its true authorship \cite{Tait}. Thus, the entropy, like the internal energy, and in contrast to the quantity of heat,  was identified as a point function which had the advantage of depending only on the instantaneous state of the body, and not on the process by which the body arrived in that state.\par
To the best of our knowledge there has never been any question of the compatibility of the first and second laws of thermodynamics. Certainly, the two laws are compatible for a classical ideal gas (ICG) because of the separability of the  temperature and volume afforded by the logarithmic function. However, this is not true, in general, for an ideal generalized gas (IGG), or a \lq quantum\rq\ gas (IQG) \cite{Landsberg}, which is the  low temperature extension of an ICG. Here, we make the distinction between an IGG and an ICG in that the former can be valid at all temperatures and not only in the high temperature limit. This means that the transition between and IGG and an ICG occurs in a manner different than taking the high temperature limit, and one in which we shall explore in the last section of this paper.\par
For an IGG, as well as an IQG, power laws are  involved that contain products of different powers of the temperature and volume. And, in fact, one finds that for all processes other than those involving pure heat conduction, the power means derived from the first and second laws are incomparable.\par
In order to rectify this incompatibility, we will prove a corollary to Carnot's theorem, asserting that the ratio of the work done in a complete cycle to the heat absorbed on expansion at constant temperature is \lq\lq the same for all substances at the same temperature\rq\rq. The corollary states that the ratio of the work done in a complete cycle to the heat absorbed on volume expansion at constant temperature is \lq\lq the same for all substances at the same volume\rq\rq. This, in effect, replaces the isotherms of the Carnot cycle by isochores without affecting the efficiency of the cycle.\par
In addition to the inverse temperature, another integrating factor for the quantity of heat will be shown to exist \cite{Einbinder}, and lead to a new adiabatic potential, which will be comparable to the entropy, and differ from it by a power of the adiabatic variable. The difference between this new potential and the entropy is that it is not a first-order homogeneous function of the volume. This will enable a comparison of means involving processes involving work where, otherwise, the first and second laws would be mute since both the internal energy and the entropy are first-order homogeneous functions of the volume for an IGG.\par
The existence of a new adiabatic potential raises the question as to the actual content and predictive power of the second law. The irresistible increase in the entropy during an irreversible process will now be confronted with the same inevitable decrease in the new adiabatic potential. Irreversibility will have to be detached from quasi-static processes which occur as a passage through a sequence of equilibrium states \cite{Born}. Irreversiblity will apply to those processes where the initial and final states differ either with respect to their temperatures and/or volumes. An  equilibration resulting in the conservation of one of the two adiabatic potentials will depend on whether the temperature ratio of the two states, or their inverse volume ratios raised to a characteristic power, is equal to the ratio of quantities of heat absorbed or rejected at these temperatures or volumes. One conserving equilibration will be necessary in order to secure a final uniform state with a common mean value. General statements can be made about such conserving equilibrations: an entropy conserving equilibration has the lowest common final mean temperature or volume, implying the maximum amount of work has been performed. In addition, thermal efficiency can never be inferior to mechanical efficiency.\par
In a companion paper \cite{Lavenda}, these statements will be translated into comparable power means, and the lack of absoluteness of these potentials as a class of equivalent means. The metrizability of such a space of equivalent means will also be studied; the notion of a metric is entirely foreign to classical thermodynamics. Probability distributions and probabilistic notions will enter naturally whenever processes inside the system occur in an  uncontrollable manner, like heat transfer and deformations. Evolution criteria will be shown to eminate from the fundamental property that power means are monotonically increasing functions of their order. There will then be shown numerous mathematical inequalities, like the Tchebychef and Jensen inequalities, all which predict an increase in entropy on the average, or to the decrease on the average of the complementary adiabatic potential. These mathematical inequalities will eliminate the need to have recourse to experiment, albeit one single experiment, to determine the sign of the entropy change, or to that of the complementary adiabatic potential.
\section{power laws}
If the absolute temperature, $T$, and volume, $V$, are chosen as the independent variables, the integrability condition for the entropy is
\begin{equation}
T\left(\frac{\partial p}{\partial T}\right)_V=\left(\frac{\partial E}{\partial V}\right)_T+p, \label{eq:int}
\end{equation}
where $p$ is the pressure, and $E$ is the internal energy. If the product $pV$ measures the absolute temperature scale then
\begin{equation}
\left(\frac{\partial E}{\partial V}\right)_T=0. \label{eq:ICG-con}
\end{equation}
In other words, in order for Boyle's law to be identical with the absolute temperature, it is necessary and sufficient that the internal energy be a \emph{linear\/} function of the temperature alone \cite{Buchdahl}.\par
If the pressure were to vary as some power of the temperature, say $T^{\alpha}$, then from (\ref{eq:int}) we would have
\begin{eqnarray}
\left(\frac{\partial E}{\partial V}\right)_T & = & \left[
\left(\frac{\partial\ln p}{\partial\ln T}\right)_V-1\right]p \nonumber\\
& = & (\alpha-1)p.\label{eq:IGG-con}
\end{eqnarray}
 Thus, the ICG condition (\ref{eq:ICG-con}) that $pV$ measure the absolute temperature, separates two domains: one in which (\ref{eq:IGG-con}) is positive, $\alpha>1$, and the attractive nature of the gas implies that it will condense, and another region in which (\ref{eq:IGG-con}) is negative, $\alpha<1$, and repulsion prevails implying that the system has a zero-point energy \cite{Einbinder}. \par
Every gas comprised of mechanically noninteracting particles obeys an equation of state of the form \cite{Einbinder}
\begin{equation}
pV=sE(V,T), \label{eq:pV}
\end{equation}
where $s>0$ is proportional to the adiabatic exponent. Introducing (\ref{eq:pV}) into (\ref{eq:int}) converts it into the differential equation
\begin{equation}
E=T\left(\frac{\partial E}{\partial T}\right)_V-\frac{V}{s}\left(\frac{\partial E}{\partial V}\right)_T. \label{eq:E-de}
\end{equation}
\par
Treating (\ref{eq:E-de}) as a Lagrange equation, the auxiliary equations are
\[\frac{dT}{T}=-\frac{s\,dV}{V}=\frac{dE}{E}.\]
There are two independent solutions, $TV^s=c$, and either $E/T=a$, or $EV^s=b$, where $a$, $b$, and $c$ are arbitrary constants. The general solution is $\Psi_1(a,c)=\Psi_1(E/T,TV^s)=0$, i.e.,
\begin{equation}
E=T\psi_1\left(TV^s\right), \label{eq:psi1}
\end{equation}
or $\Psi_2(b,c)=\Psi_2(EV^s,TV^s)=0$, i.e.,
\begin{equation}
E=V^{-s}\psi_2\left(TV^s\right), \label{eq:psi2}
\end{equation}
where $\Psi_i$ and $\psi_i$ are arbitrary functions. The coefficients of $E$, namely, $1/T$ and $V^s$, will later be appreciated as integrating factors for the quantity of heat.\par The  functions $\psi_i$ are solutions to the adiabatic equation
\begin{equation}
T\left(\frac{\partial\psi_i}{\partial T}\right)_V-\frac{V}{s}\left(\frac{\partial\psi_i}
{\partial V}\right)_T=0. \label{eq:adiabatic}
\end{equation}
The auxiliary equations to (\ref{eq:adiabatic}) are
\[\frac{dT}{T}=-\frac{s\,dV}{V}=\frac{d\psi_i}{0}.\]
There are again two independent solutions, $\psi_i=a$ and $TV^s=b$, so that the general solution to (\ref{eq:adiabatic}) is $\psi_i=\psi_i(z)$, in which $T$ and $V^s$ appear only through the combination $z=TV^s$.\par
For $\psi_1=\mbox{const}$., (\ref{eq:psi1}) is the thermal equation of state for an ICG, while for $\psi_2=\mbox{const}$., (\ref{eq:psi2}) is the zero-point energy. Whereas $E=aT$ is an approximate relation, valid in the high temperature limit, so too  $E=bV^{-s}$  can be considered an approximate relation, this time valid in the low temperature limit \cite{Einbinder}. These are the extreme cases where the internal energy is a function of either the absolute temperature, in the high temperature limit, or of the volume, in the low temperature limit. The zero-point expresses the fact that particle interactions are repulsive, $(\partial E/\partial V)_T<0$, and such a system would be entirely mechanical since $dQ=dE+p\,dV=0$. However, there is another possibility in which the internal energy, as well as the entropy, tends to zero monotonically with the temperature.\par
The thermal equation of state (\ref{eq:psi1}) can be considered as an IGG, i.e., one for which the particle number, $N=\psi_1(V,T)$, is variable, being a function of the temperature. This will provide a very profound analogy between a two phase classical system, like a Carnot engine, that was analyzed by Clapeyron by an equation bearing his name, and an IGG which does not conserve the number of particles.
\par
In the case that $E$ tends to zero monotonically with $T$, $\psi_2(z)$ may be approximated by $L(z)=cT^{\alpha}V^{s\alpha}$ at low temperatures where all we demand for the present is that $\alpha>0$.  According to (\ref{eq:pV}), the internal energy, $E=cT^{\alpha}V^{(\alpha-1)s}$, gives a pressure $p=scT^{\alpha}V^{(\alpha-1)s-1}$. For dynamic stability we require
\[\left(\frac{\partial p}{\partial V}\right)_T=[(\alpha-1)s-1]\frac{p}{V}<0.\]
\par
However, if the internal energy is to retain its property of being a first-order homogeneous function, we must have $(\alpha-1)s=1$, implying 
\begin{equation}
\left(\frac{\partial p}{\partial V}\right)_T=0, \label{eq:phase}
\end{equation} or a phase equilibrium \cite{Einbinder}. Varying the volume at constant temperature leaves the vapor pressure constant by having the liquid either evaporate or condense. This is precisely the condition under which the Clapeyron equation is valid. For an IGG, the volume of the second phase is nil since there is no longer particle conservation, $N=\psi_1(V,T)\neq\mbox{const}$. This fine balance keeps the pressure independent of the volume and the internal energy a first-order homogeneous finction
\begin{equation}
p=scT^{q/r}\;\;\;\;\;\;\;\;\;\;\;\; E=cT^{q/r}V. \label{eq:p&E}
\end{equation}
\par
In the low temperature limit  $\psi_2(z)$ can be replaced by 
\begin{equation}
L(z)=a+cz^{q/r}, \label{eq:L}
\end{equation} 
where $a>0$.  Then, since $EV^s=L(z)$ 
\[E=V^{-s}\left(a+cz^{q/r}\right),\]
and $(\partial E/\partial V)_T<0$ if $a$ is finite, or $(\partial E/\partial V)_T>0$ if $a$ vanishes, and $q>r$. In the later case, the pressure must satisfy (\ref{eq:phase}) so that $s=r/(q-r)$. Furthermore, $E/T=\psi_1(z)=L(z)/z$, and $d\left(EV^s\right)/z=dL(z)/z=:dS$, where
\begin{equation}
S(z)=\frac{q}{q-r}cz^{1/s}=\frac{q}{q-r}\psi_1(z) \label{eq:S}
\end{equation}
is the entropy.\par Hence, no decision can be made between the Planck ($S=0$ at $T=0$) and the Nernst $(S=\mbox{const.}$ at $T=0$) formulations of the third law because the concept of absolute entropy is meaningless. Moreover, we will see in the following paper that an absolute $L(z)$ is also meaningless since only its difference is measurable.\par
The difference between the enthalphy,
\[
H=U+pV=(1+s)cT^{q/r}V, \]
and $T$ times the entropy, (\ref{eq:S}), is 
\begin{equation}
G=\left(1+s-\frac{q}{q-r}\right)cT^{q/r}V. \label{eq:G}
\end{equation}
The Gibbs free energy (\ref{eq:G}) presents itself as a measure of non-extensivity, and vanishes when $s=r/(q-r)$ \cite{Einbinder}. 
\par
\section{the second laws}
The quantity of heat, $dQ$, which must be absorbed by a body to makes its temperature rise to $T+dT$ and its volume expand to $V+dV$ is
\begin{equation}
dQ=M\,dV+N\,dT. \label{eq:caloric}
\end{equation}
This caloric equation, familiar to all the early thermodynamicists, undoubtedly fell out of favor due to the fact \cite{Clausius}
\begin{quote}
that $Q$ cannot be a function of $V$ and $T$, if these variables are independent of each other. For if it were, then by the well-known law of the differential calculus, that if a function of two variables is differentiated with respect to both of them, the order of differentiation is indifferent
\end{quote}
and this was definitely not so with (\ref{eq:caloric}). In fact, the exactness condition of the internal energy,
\[dU=dQ-p\,dV=(M-p)\,dV+N\,dT,\]
shows that their difference gave \cite{Thomson}
\begin{equation}
\frac{dp}{dT}=\frac{\partial M}{\partial T}-\frac{\partial N}{\partial V}. \label{eq:I}
\end{equation}
\par
This led Planck \cite{Planck} to remark that the notation $dQ$
\begin{quote}
has frequently given rise to misunderstanding, for $dQ$ has been repeatedly regarded as the differential of a known finite quantity $Q$. This faulty reasoning may be illustrated by the following example.
\end{quote}
And the example Planck gave resulted in (\ref{eq:I}), which Thomson took merely as a statement of the first law.\par The statement that he took as the second law was Clapeyron's equation
\begin{equation}
\frac{dp}{dT}=\frac{M}{C(T)}, \label{eq:Clapeyron}
\end{equation}
where $C(T)$ is the reciprocal of the Carnot function, which Thomson showed was equal to the absolute temperature, $T$. That is, the ratio of the total work done in an infinitesimal cycle, $\frac{dp}{dT}\,dTdV$, to the ratio of the heat absorbed in the first branch of the Carnot cycle, $M\,dV$,
\begin{equation}
\frac{\frac{dP}{dT}\,dT}{M}=\frac{dT}{C(T)} \label{eq:Carnot-T}
\end{equation}
must be the product of $dT$ and a function of $T$ only. The \lq\lq very remarkable theorem that $dp/dT\big/M$ must be the same for all substances at the same temperature was first given (although not in precisely the same terms) by Carnot\rq\rq \cite{Thomson}. \par
For an ICG, $M=p$, because the internal energy is a function of the absolute temperature alone, and $N$, it was realized by Clausius, \lq\lq can be a function of $T$ only. It is even probable that this magnitude [$N$], which represents the specific heat of the gas at constant volume, is a constant.\rq\rq\par
In constrast, for an IGG,
\begin{equation}
M=\left(1+\frac{1}{s}\right)cT^{q/r}=\frac{q}{r}p, \label{eq:M}
\end{equation}
and
\begin{equation}
sN=\frac{q}{r}cT^{1/s}V. \label{eq:N}
\end{equation}
Expression (\ref{eq:M}) invalidates Clausius's conclusion that \lq\lq a permanent gas, when expanded at constant temperature, takes up only so much heat as is consumed doing external work during the expansion\rq\rq. Clausius's conclusion is based on the fact that the working substance was an ICG, obeying $pV=T$, while (\ref{eq:M}) shows that an IGG absorbs more since $q>r$.\par
To convert $dQ$ into the total differential of a certain function, we introduce the integrating factor $\lambda$, and require
\begin{equation}
\frac{\partial\lambda M}{\partial T}=\frac{\partial\lambda N}{\partial V}. \label{eq:lambda}
\end{equation}
The exactness condition (\ref{eq:lambda}) can be rearranged to read
\begin{equation}
\left(\frac{\partial M}{\partial T}-\frac{\partial N}{\partial V}\right)=-M\frac{\partial\ln\lambda}{\partial T}+N\frac{\partial\ln\lambda}{\partial V}.
\label{eq:exact}
\end{equation}
\par
By a composite system argument, $\lambda$ can only be a function of $T$ or $V$. For consider two simple fluids in thermal contact; such a system will have three independent variables $T$, $V_1$ and $V_2$. The integrating factor can only be a function of the common variable $T$ \cite{Caratheodory}. Hence,
\[\frac{\left(\frac{\partial M}{\partial T}-\frac{\partial N}{\partial V}\right)}
{M}=\frac{d\ln\lambda}{dT}.\]
On the strength of the exactness condition for the internal energy (\ref{eq:I}), this is equivalent to the Clapeyron equation (\ref{eq:Carnot-T}), and on the strength of Carnot's theorem $C(T)=T$ is the integrating denominator for the increment in the heat, $dQ$, giving the entropy, (\ref{eq:S}), which depends on $V$ and $T$ only through the combination $z=TV^s$.\par
Now consider two simple fluids in mechanical contact, for which the independent variables are $V$, $T_1$, and $T_2$. Again, the integrating factor $\lambda$ can only be a function of the variable in common to both subsystems so that the exactness condition (\ref{eq:exact}) now reduces to
\begin{equation}
\frac{\frac{dp}{dT}\,dT\,dV}{N\,dT}=d\ln\lambda(V). \label{eq:Carnot-V}
\end{equation}
By interchanging isochores for isotherms, $N\,dT$ is the quantity of heat absorbed in the first segment of an equivalent Carnot cycle, as we shall discuss in the next section. The very remarkable fact, equivalent to Carnot's theorem, is that the right-hand side is the product of $dV$ and a function only of the volume. Thus, \emph{$dp/dT/N$ is the same for all substances at the same volume\/}. For an IGG, $\lambda(V)=V^s$ \cite{Einbinder}, which reduces to $\lambda(V)=V^{R/C_v}$ for an ICG, where $C_v$ is the heat capacity at constant volume, and $R$ is the gas constant, which will only be introduced in conjunction with $C_v$.\par The integrating factor $\lambda(V)$ for the increment in the heat, $dQ$, now gives the point function (\ref{eq:L}),
which, again, depends on $V$ and $T$ only through the combination, $z$. However, whereas the entropy (\ref{eq:S}) is extensive, the potential  (\ref{eq:L}) is not. \par
Consider any two states $1$ and $2$, with $T_1>T_2$ for concreteness. Any process connecting the two states will be said to be irreversible if
\begin{equation}
z_1>z_2. \label{eq:z}
\end{equation}
Rearranging (\ref{eq:z}) we get $V_1^s/V_2^s>T_2/T_1$ so that the thermal efficiency,
\[
\eta_t=1-\frac{T_2}{T_1}\ge 1-\frac{V_1^s}{V_2^s}=\eta_v, \]
can never be inferior to the mechanical efficiency, $\eta_v$.\par
If $Q_1$ and $Q_2$ are quantities of heat absorbed and rejected at $T_1$ and $T_2$, respectively, then we say that there is \emph{thermal equilibration\/} if
\[
\frac{T_2}{T_1}=\frac{Q_2}{Q_1}\ge\frac{V_1^s}{V_2^s}, \]
whereas there is \emph{mechanical equilibration\/} if
\[
\frac{V_1^s}{V_2^s}=\frac{Q_2}{Q_1}\le\frac{T_2}{T_1}. \]
More work can be accomplished during a thermal equilibration than a mechanical one since the system achieves a lower final temperature, $T_2$.\par
For any infinitesimal narrow Carnot cycle, the ratio $Q_2/Q_1$ may be replaced by its differential, $dQ_2/dQ_1$. Thermal equilibration results when
\begin{equation}
\oint\frac{dQ}{T}=0,\label{eq:I-1}
\end{equation}
and
\begin{equation}
\oint\,V^s\,dQ\ge0,\label{eq:I-2}
\end{equation}
while mechanical equilibration requires
\begin{equation}
\oint\,V^s\,dQ=0,\label{eq:II-1}
\end{equation}
and
\begin{equation}
\oint\frac{dQ}{T}\le0,\label{eq:II-2}
\end{equation}
since any irreversible cycle is necessarily less efficient than a Carnot cycle, $dQ_2/dQ_1<T_2/T_1$.\par For processes of pure thermal conduction, we would identify (\ref{eq:II-1}) with the first law, and the conservation of energy over the entire cycle, and (\ref{eq:II-2}) as the statement of the second law. However, for purely mechanical interactions neither (\ref{eq:II-2}) nor the first law would give an evolutionary criterion since the internal energy and entropy are first-order homogeneous functions of the volume. The criteria of mechanical equilibration would fix the final volume as the weighted arithmetic mean of the partial volumes, according to (\ref{eq:I-2}), while (\ref{eq:I-1}) would show that $L$ would tend to decrease since the arithmetic mean is inferior to the power mean of order $q/(q-r)$.\par
If we split the cycle up into two segments, $A\rightarrow B$, and $B\rightarrow A$, where the former contains all the irreversibility, then (\ref{eq:I-2}) gives
\[\int_A^B V^s\,dQ>L_B-L_A,\]
while (\ref{eq:II-2}) becomes
\[S_B-S_A>\int_A^B\frac{dQ}{T}.\]
If the infinitesimal increment in the heat is the sum of the infinitesimal heat $dQ_e$ introduced into the system, or extracted from it, and the sum of irreversible heat transfers within the system,
\[dQ=dQ_e+\sumi dQ_i,\]
then for an isolated  system $dQ_e=0$,
\begin{equation}
L_B-L_A<\sumi\int_A^B V^sdQ_i,\label{eq:Clausius-bis}
\end{equation}
under the condition that $\sumi\oint dQ_i/T=0$, or
\begin{equation}S_B-S_A>\sumi\int_A^B\frac{dQ_i}{T},\label{eq:Clausius}
\end{equation}
under the condition $\sumi\oint V^sdQ_i=0$. Inequality (\ref{eq:Clausius}) is referred to as Clausius's inequality, while inequality (\ref{eq:Clausius-bis}) appears to be novel.\par
For each individual process of heat transfer, $dQ_i$ will appear twice: once as a positive quantity and once as a negative quantity. The second law (\ref{eq:Clausius}) asserts that the integrating denominator in the former is smaller than in the latter, since \lq heat flows spontaneously from a hotter to a colder body at constant volume.\rq\  Alternatively, according to (\ref{eq:Clausius-bis}),  \lq as heat is absorbed  the gas  expands at constant temperature\rq, so that the integrating factor of the former will be smaller than the latter.\par
Either (\ref{eq:Clausius-bis}) or (\ref{eq:Clausius}) can be taken as statements of the second law, depending on the constraints imposed. It is quite remarkable that both criteria can be formulated in terms of \emph{comparable\/} means, and their fundamental property that the power mean is a monotonic increasing function of its order will  be  investigated thoroughly in the following paper.
\par 
\section{An equivalent Carnot cycle}
An equivalent cycle to that of Carnot can be obtained by replacing isothermals by isochores. This will prove a very remarkable theorem that $dp/dT\big/N$ must be the same for all substances at the same volume, which is precisely analogous to Carnot's theorem that $dp/dT\big/M$ must be the same for all substances at the same temperature.\par
The cycle consists of:
\begin{enumerate}
\item Absorption of a quantity of heat $Q_1$, at a constant volume $V_1$, by compression which raises the absolute temperature from $T_1$ to $T_2$, and, consequently, increases the pressure.
\item An adiabatic expansion to a state of larger volume $V_2$, and lower temperature $T_3$.
\item The rejection of a quantity of heat $Q_2$, by expansion which lowers the temperature to $T_4$ at constant volume $V_2$.
\item An adiabatic compression which restores the system to volume $V_1$, and temperature $T_1$.
\end{enumerate}
The adiabatic branches provide the following ratios
\[\frac{V_1^s}{V_2^s}=\frac{T_3}{T_2}=\frac{T_4}{T_1}.\]
The ratio of the heat rejected, $|Q_{3\rightarrow 4}|$, to the heat absorbed,
$Q_{1\rightarrow 2}$ is
\[\frac{|Q_{3\rightarrow 4}|}{Q_{1\rightarrow 2}}=\frac{V_1^s}{V_2^s},\]
and the efficiency of the engine
\[\eta_v=1-\frac{V_1^s}{V_2^s},\]
is the same as the Carnot, thermal, efficiency, $\eta_t$.\par
Now, for an infinitesimal cycle, the ratio of the total work performed to the heat absorbed at constant volume is (\ref{eq:Carnot-V}), which is the same for all substances at the same volume. This, as we have shown in the last section, is equivalent to the Carnot theorem (\ref{eq:Carnot-T}).\par Both (\ref{eq:Carnot-V}) and (\ref{eq:Carnot-T}) gives the same Clapeyron equation, which, when integrated, gives the pressure in (\ref{eq:p&E}), independent of volume for a homogeneous system. And while  it was realized that $dp/dT$ was a function of the temperature alone \cite{Clausius}, it was not appreciated that the mechanical ICG equation of state, $pV=T$, could not be used to evaluate the work.

\section{The transition IGG$\rightarrow$ICG}
The absolute temperature, $T$, and the empirical temperature, $t$, will coincide only for an ICG. For an ICG, $pV$ reads the temperature, and $E$ is a function of the temperature alone, independent of the volume. In contrast for an IGG, $p$ will be independent of the volume, and $E$ will be a function of it. In addition, $E$ will no longer be a linear function of the temperature.\par
The only demand made by the zeroth law is that when two identical systems are placed in thermal contact their empirical temperatures be the same when a state of mutual thermal equilibrium has been reached. Once the empirical scale has been chosen, the absolute temperature must be a monotonically increasing function of it, viz.,
\[T(t)=t^r,\]
with $r\ge1$.\par On the empirical temperature scale, $E$ will vary as $t^q$, where $q\ge r$ with the equality sign pertaining to the ICG limit. Whereas the exponent $r$ is related to the average kinetic energy of the particles, the exponent $q$ is related to additional forms of energy, like the energy required to create or annihilate particles, since for an IGG the particle number is not conserved. As a matter of fact, the vanishing of the difference of the two exponents will signal the transition from an IGG to an ICG, as we shall now show. This is in distinction to the transition between an IQG and an ICG which occurs only in the high temperature limit.\par
It is notable that in the limit as $q\rightarrow r$, the entropy (\ref{eq:S}) is of the indeterminate form $0/0$. Applying L'H\^{o}pital's rule, we get
\[\lim_{q\rightarrow r}S(z)=c\ln z,\]
where $z=TV^{R/C_v}$. Identifying $c$ as $C_v$ gives
\[\lim_{q\rightarrow r}S(z)=C_v\ln T+R\ln V.\]\par
In the same limit,
\[\lim_{q\rightarrow r}L(z)=C_v T^{R/C_v},\]
and consequently, either
\[dE=T\,dS(z)-p\,dV=C_v\,dT\]
or
\[dE=\frac{1}{V^{R/C_v}}dL(z)-p\,dV=C_v\,dT,\]
which is the thermal equation of state of an ICG.

\end{document}